\documentstyle[12pt]{article}

\def\+{{+\!\!\!+}}

\def\d{\partial}

\def\l{\lambda}

\def\s{\sigma}

\def\pmb#1{\setbox0=\hbox{#1}%
\kern.0em\copy0\kern-\wd0
\kern-.04em\copy0\kern-\wd0
\kern.08em\copy0\kern-\wd0
\kern-.04em\raise.0433em\box0 }         %poor man's bold macro (TexBook)
\newcommand{\nc}{\newcommand}
\nc{\beq}{\begin{equation}}
\nc{\eeq}[1]{\label{#1}\end{equation}}
\nc{\ber}{\begin{eqnarray}}
\nc{\eer}[1]{\label{#1}\end{eqnarray}}
\nc{\pek}[1]{\cite{#1}}
\nc{\enr}[1]{(\ref{#1})}
\nc{\kal}[1]{{\cal{#1}}}
\nc{\dott}{\;\cdot\;}
\begin{document}
\newcommand{\inv}[1]{{#1}^{-1}} %inverse
\renewcommand{\theequation}{\thesection.\arabic{equation}}
\newcommand{\be}{\begin{equation}}
\newcommand{\ee}{\end{equation}}
\newcommand{\bea}{\begin{eqnarray}}
\newcommand{\eea}{\end{eqnarray}}
\newcommand{\re}[1]{(\ref{#1})}
\newcommand{\qv}{\quad ,}
\newcommand{\qp}{\quad .}
%\begin{titlepage}
%\title{}
\begin{center}

                                \hfill    USITP-00-10\\
                                \hfill    hep-th/0009113\\

\vskip .3in \noindent

\vskip .1in

{\large \bf {Relevant boundary perturbations of CFT: A case study}}
\vskip .2in

{\bf Tasneem Zehra Husain}\footnote{tasneem@physto.se}
 and {\bf Maxim Zabzine}\footnote{zabzin@physto.se}  \\

\vskip .15in

\vskip .15in

\vskip .15in

 {\em Institute of Theoretical Physics,
University of Stockholm \\
Box 6730,
S-113 85 Stockholm SWEDEN} \\
\bigskip

\vskip .15in

\vskip .1in
\end{center}
\vskip .4in
\begin{center} {\bf ABSTRACT }
\end{center}
\begin{quotation}\noindent
 We consider simple CFT models which contain massless bosons, massless 
 fermions or a supersymmetric combination of the two, on the strip. We
study 
 the deformations of these models by relevant boundary operators. In
particular,
 we work out the details for a boundary operator with a quadratic
dependence 
 on the fields and argue that some of our results can be extended to a
more
 general situation. In the fermionic models, several subtleties arise due
to 
 a doubling of zero modes at the UV fixed point and a ``GSO projected''
RG 
 flow. We attempt to resolve these issues and to discuss how bulk
symmetries 
 are realised along the flow. We end with some speculations on possible
string 
 theory applications of these results.  
\end{quotation}
\vfill
\eject

\section{Introduction and motivation}

 Boundary conformal field theories (BCFT) find applications in vastly 
 different branches of physics ranging from condensed matter systems to 
 string theory. 
 
 BCFTs can be perturbed by an operator $\Phi$ which couples only to the 
 boundary. Such a perturbation leads to new boundary conditions which
 may even break conformal invariance. If the operator $\Phi$ has conformal 
 dimension $h=1$, it is possible to stay at the conformal point 
 to all orders in the coupling constant. However if $h\neq 1$, the
boundary 
 perturbation introduces a length scale into the theory and there is hence
a 
 RG-flow. When $h>1$ the perturbation is irrelevant and the flow keeps us
 within the same BCFT; for $h<1$ the perturbation is 
 relevant and the BCFT we flow to in the IR is different to the one we
started 
 out with. In general it is a difficult problem to find the IR BCFT 
 corresponding to a given relevant perturbation, nevertheless, several 
 non-trivial examples have been worked out using the thermodynamic Bethe 
 ansatz \cite{Fendley:1994rh} and perturbative techniques 
 \cite{Fendley:1994ms}.
 The study of relevant boundary perturbations of BCFTs is 
 mainly motivated by quantum impurity problems in condensed matter
 theory \cite{Saleur:1998hq} and by tachyonic condensation in string 
 theory \cite{Harvey:2000na}. 

 In the present note we would like to study relevant boundary
perturbations of 
 a few simple BCFTs. In contrast to previous works, which were based on 
 integrable model or perturbative techniques, we use the canonical 
 quantization 
 aproach and concentrate on possible algebraic structures that may arise. 
 We consider several free theories which are solvable and can be regarded
as 
 a reasonable first approximation to interacting theories. We argue that
some 
 results obtained for free theories can be extended to general 
 interacting theories as well. 
 
 Our main motivation for this work comes from string theory applications 
 (i.e., the RNS string model in a tachyonic background field
\cite{Harvey:2000na}, 
 \cite{Husain:2000wa}). However we hope that some of our results might
also applied 
 to other branches of physics such as impurity problems. Therefore,
throughout 
 the paper we try to avoid any direct references to string theory,
reserving 
 such comments/discussions for the end. 
 
 We define our theory, which is conformally invariant in the bulk, on 
 a strip $R\times[0,L]$ with two spatial boundaries. In general, both  
 boundaries can have different boundary interactions. Though we work with
 Minkowski signature, there are no problems in principle, in going to the 
 Euclidean version and considering the same kinds of theories on a disk, 
 an annulus and etc. In all cases, this generalization is straightforward 
 and throughout the paper, we comment on the Euclidean versions of the 
 models we consider.

 The paper is organized as follows: In section 2 we consider a theory with
a 
 massless free boson living in the bulk and a potential coupled to the 
 boundary. We argue that for a general potential, the chiral and
anti-chiral 
 algebras survive, and infact become non-trivially related to each other 
 through the boundary perturbation. We work out these algebras
in 
 detail, assuming a quadratic form for the boundary perturbation. 
 In section 3 we consider the fermionic couterpart of the previous model. 
 It turns out that the construction of relevant boundary perturbations 
 for these models involves some subtleties due to appearance of extra zero 
 modes. Some general statements about such 'extra zero modes' have
previously
 been made in the string theory literature, \cite{Witten:1998cd} and
 \cite{Harvey:2000na}. We try, in the framework of our model, to 
 present a detailed discussion of the mechanism whereby these zero modes
arise.
  
 Using the above results, we construct a supersymetric 
 model in section 4. We are able to do this only for a quadratic boundary 
 potential and we offer some reasons as why it might be difficult to do so 
 in general. In section 5 we go back to the bosonic and fermionic models
but 
 this time, with different perturbations on each of the two boundaries;
 we discuss the realization of bulk symmetries in this case.
 In section 6 we speculate on the string theory applications
 of our results. The construction of the RNS
 string model in a tachyonic background field seems to be quite
restrictive 
 and we have been able handle only a quadratic tachyonic field. This could 
 be a sign that possible tachyonic backgrounds are limited by 
 self-consistency of the theory, but ofcourse, more work needs to be done 
 before one can reach any final conclusions. We also propose directions
for 
 further study. 

\section{Free boson theory}

In this section we study a simple BCFT perturbed by a relevant 
boundary operator. We consider the theory on a strip with the 
following Lagrangian 
\be
\label{A.1}
L= \frac{1}{2} \int\limits_{0}^{L} d\s\, \d_\alpha \phi \d^\alpha \phi + 
 \lambda V(\phi(0)) - \lambda V(\phi(L)) ,
\ee
 and Minkowski signature $(1,-1)$. The Lagrangian (\ref{A.1}) gives
 rise to the following equation of motion and boundary conditions
\be
\label{A.2}
 (\d^2_\tau -\d^2_\s)\phi=0,\,\,\,\,\,\,\,\,\,\,\,\,\,\,\,\,
( \phi'+\lambda \d_\phi V(\phi))|_{0,L}=0.
\ee
 The boundary conditions have the same form on both boundaries, however
the 
 model is not parity invariant. A parity transformation $\Omega$
 ($\sigma\,\, \rightarrow\,\, L-\sigma$) is in fact equivalent to 
 changing the sign of the coupling constant $\lambda$. 
 The equation of motion (\ref{A.2}) is satisfied if $\phi(\tau,\sigma)$ 
 is the sum of two arbitrary functions of $(\tau+\sigma)$ and
$(\tau-\sigma)$.
 
 Introducing the notation $z=\exp(-i
 \frac{\pi}{L}(\tau+\s))$ 
and $\bar{z} = \exp(-i\frac{\pi}{L}(\tau-\s))$ we can define the 
 following currents 
\be
\label{A.3}
 J(z) = \sqrt{\frac{2}{L}}\d_{+} \phi = \sum\limits_{n} J_n z^{n},
 \,\,\,\,\,\,\,\,\,\,
 \bar{J}(\bar{z}) =\sqrt{\frac{2}{L}} \d_{-} \phi =  \sum\limits_{n} 
 \bar{J}_n \bar{z}^{n}.
\ee
 where $\partial_{\pm}=\partial_\tau \pm \partial_\s$. 
 The Wick rotation $\tau \rightarrow it$ makes $J(z)$ holomorphic and 
 $\bar{J}(\bar{z})$ anti-holomorphic. 
 Canonical commutation relations imply that these currents satisfy the 
 following chiral and antichiral algebras
\be
\label{A.4}
 [J_n, J_m] = n\delta_{n+m},\,\,\,\,\,\,\,\,\,\,\,\,
 [\bar{J}_n, \bar{J}_m] = n\delta_{n+m}.
\ee 
 
 The boundary conditions in (\ref{A.2}) lead to a nontrivial relation 
 between $J_n$ and $\bar{J}_n$. However, this relation should be such that 
 it respects the algebra (\ref{A.4}). In other words, holomorphic and 
 anti-holomorphic currents are linked by a unitary transformation in the
 corresponding Fock space.
 
 Using the Sugawara construction we can write down generators 
\be
\label{A.5}
L_k = \frac{1}{2} \sum\limits_{n} J_{n}
J_{k-n},\,\,\,\,\,\,\,\,\,\,\,\,\,\,
\bar{L}_k = \frac{1}{2} \sum\limits_{n} \bar{J}_{n} \bar{J}_{k-n}
\ee
 which obey the standard Virasoro algebras. 
 Note that $L_k \neq \bar{L}_k$. This is as it should be, since the
Lagrangian 
 (\ref{A.1}) is not invariant under conformal transformations.

We now want to work out a simple example, using a specific form for 
the boundary perturbation. Let us take $V(\phi)=\frac{1}{2}\phi^2$ which 
makes the boundary conditions linear
\be
\label{A.6}
 (\phi'+\lambda \phi)|_{0,L}=0. 
\ee
 Equation (\ref{A.2}) subject to boundary conditions (\ref{A.6}) can
 be reduced to a Sturm-Liouville problem. The properly normalized solution
is 
 given by
\be
\label{A.7}
\phi(\tau,\sigma) = \frac{\sqrt{2L}}{\pi} \sum\limits_{n \neq 0}
 \frac{e^{-in\frac{\pi}{L} \tau}}{\sqrt{n^2 + \tilde{\lambda}^2}}
 a_n [\cos(n\frac{\pi}{L}\sigma) - \frac{\tilde{\lambda}}{n}
 \sin(n\frac{\pi}{L}\sigma)]
\ee
 where $\tilde{\lambda}= \frac{L\lambda}{\pi}$ is the dimensionless
coupling
 constant. At $\lambda = 0$, due to the zero modes that arise, the 
 above solution contains an extra term, $a_0 \tau$. 
Using 
 completeness of eigenfunctions of the
 corresponding Sturm-Liouville problem, one can find the 
 commutation relations $[a_n, a_m] = n\delta_{n+m}$. Reality
 of $\phi$ implies that $a_n^{+}= a_{-n}$. 
 
The components of chiral and antichiral currents are now given by
\be
\label{A.8}
 J_n = - \frac{\tilde{\l}+in}{\sqrt{n^2+\tilde{\l}^2}} a_n,
\,\,\,\,\,\,\,\,\,\,\,\,\,
 \bar{J}_n =  \frac{\tilde{\l}-in}{\sqrt{n^2+\tilde{\l}^2}} a_n.
\ee 
 and are related by the following transformation
\be
\label{A.9}
 J_n - e^{i\varphi(n,\tilde{\l})} \bar{J}_n = 0,
\,\,\,\,\,\,\,\,\,\,\,\,\,\, e^{i\varphi(n,\tilde{\l})}=
 \frac{in+\tilde{\l}}{in-\tilde{\l}}.
\ee
 
 Altogether (\ref{A.4}) and (\ref{A.9}) imply the following algebra
\be
\label{A.10}
[J_n, J_m] = n\delta_{n+m},\,\,\,\,\,\,\,\,\,\,\,\,
 [\bar{J}_n, \bar{J}_m] = n\delta_{n+m},\,\,\,\,\,\,\,\,\,\,\,\,
 [J_n, \bar{J}_m] = e^{i\varphi(n,\tilde{\l})} n\delta_{n+m}. 
\ee
 The above algebra is all we need
 to calculate Green's functions of chiral and antichiral currents
 (or other objects constructed from them). From the algebra we
 can see that the correlators 
\be
\label{A.11}
\langle J(z_1) J(z_2)... J(z_k)\rangle,\,\,\,\,\,\,\,\,\,\,\,\,\,
\langle \bar{J}(\bar{z}_1) \bar{J}(\bar{z}_2)... \bar{J}(\bar{z}_k)\rangle
\ee
 are exactly the same as in the unperturbed conformal field theory.
 However the mixed correlators
\be
\label{A.12}
\langle J(z_1) J(z_2)... J(z_p)\bar{J}(\bar{z}_{p+1}) 
\bar{J}(\bar{z}_{p+2})... \bar{J}(\bar{z}_k)\rangle
\ee 
 are different because of the last commutation relation in (\ref{A.10}).
 It is easy to convince oneself that this statement is rather generic for
 relevant boundary perturbations, though the explicit form of the third 
 commutation relation may vary. Certainly, the relation
 between $J_n$ and $\bar{J}_n$ will not always be linear. 
 In general, a representation of the Virasoro algebra is labelled by 
 the eigenvalue of the $a_0$ operator. However, along the flow, there is
no 
 $J_0$ (i.e no zero mode $a_0$), and hence there is one unique
representation. 

 Using the definitions (\ref{A.5}) one can obtain left and right Virasoro
 generators that obey the required Virasoro algebras independently.
 However $L_k \neq \bar{L}_k$ except in the limiting values of the
coupling
 constant, $\tilde{\l}=0$ and $\tilde{\l}=\infty$. As can be seen from
 (\ref{A.9})

$$ \lim\limits_{\tilde{\l}\rightarrow 0} e^{i\varphi(n,\tilde{\l})} = 1
\,\,\,\,\,\,
\Rightarrow \;\;\;\;\;\; J_n - \bar{J}_n = 0 
\,\,\,\,\,\, {\rm at} 
\,\,\,\,\,\, \tilde{\l} = 0, $$
\be
\label{A.13}
 \lim\limits_{\tilde{\l}\rightarrow \infty} e^{i\varphi(n,\tilde{\l})} =
-1
\,\,\,\,\,\,
\Rightarrow \;\;\;\;\;\; J_n + \bar{J}_n = 0 
\,\,\,\,\,\, {\rm at} 
\,\,\,\,\,\, \tilde{\l} = \infty. 
\ee 

The above relations between $J_n$ and $\bar{J}_n$ imply that we start out 
with Neumann boundary conditions in the UV and flow to Dirichlet
conditions 
in the IR.  
   
\section{Free fermion theory}

 We would now like to construct the fermionic counterpart of the
 previous example. Since we want to eventually combine bosonic and 
 fermionic theories to form a single supersymmetric model, we simply 
 obtain the fermionic theory by supersymmetrising the bosonic one. 

 Using the supersymmetric transformation 
\be
\label{B.14}
 \delta \phi = - \epsilon^{+} \psi_{+} - \epsilon^{-} \psi_{-} ,
\ee
on equation (\ref{A.2}), we get the fermionic boundary conditions 
\be
\label{B.15}
[ \partial_\s (\eta \psi_{+} + \psi_{-}) + 
 \l \partial_\phi^2V(\phi) (\eta\psi_{+} + \psi_{-})]|_{0,L}=0 ,
\ee
 where $\epsilon^+=\eta \epsilon^-$, for $\eta = \pm 1$, and $\psi_{\pm}$
are
components of a Majorana spinor .  
The relative sign of $\eta$ determines whether we are in the Neveu-Schwarz 
($\eta_0 = - \eta_L$) or Ramond ($\eta_0 = \eta_L$) sector.

In analogy to the bosonic case, we would expect that the boundary
condition
(\ref{B.15}) iterpolates between Neumann and Dirichlet conditions
at $\l =0$ and $\l=\infty$ respectively. This is almost, but not quite,
true. 
The complication arises due to having a dimensionful coupling constant 
$\l$ in the boundary condition (\ref{B.15}); to make the dimensions come
out
right, we must also include a derivative on the fermions. 
Thus, in the UV limit ($\l =0$) we actually have
$\partial_\s (\eta \psi_{+} + \psi_{-})=0$ rather than the 
standard Neumann condition $\eta \psi_{+} - \psi_{-}=0$. For 
nonzero modes these conditions are completely equivalent. However
subtleties arise when we consider the zero modes. 
Our 'modified' Neumann condition contains a derivative and thus does not 
impose any constraints on the zero modes unlike the standard Neumann case
which reduces the number of zero modes from two to one. 

The boundary condition (\ref{B.15}) then, can be seen to describe a flow 
along which the number of zero modes (or alternately, the degeneracy of
the 
ground state) decreases. Also, the UV fixed point of this 
flow does not coincide with that of the standard fermionic model precisely 
because of this 'doubling' of zero modes. 
 
We now analyze the fermionic model in detail.
Since we are perturbing only by a boundary operator, the bulk physics
remains
unchanged and we have the usual equations of motion
\be
\label{B.16}
 \partial_{+} \psi_{-} = 0,\,\,\,\,\,\,\,\,\,\,\,\,\,\,\,\,\,\,\,\,
 \partial_{-} \psi_{+} = 0.
\ee 
Taking into account (\ref{B.16}) the conditions (\ref{B.15}) can be
rewritten as follows
\be
\label{B.17}
[ \partial_\tau (\eta \psi_{+} - \psi_{-}) + 
 \l \partial_\phi^2V(\phi) (\eta\psi_{+} + \psi_{-})]|_{0,L}=0.  
\ee 
 For the sake of simplicity we take $V(\phi)=\frac{1}{2}\phi^2$.  
 
We now construct the action which gives rise to
the above equations of motions and boundary conditions. Since 
in writing down the action we would like to avoid having extra derivatives
on 
fermions, we introduce an auxiliary fermion $d$ on the boundary, 
as follows
\be
\label{B.18}
[ \partial_\tau d + 
  \sqrt{\l}  (\eta\psi_{+} + \psi_{-})]|_{0,L}=0,\,\,\,\,\,\,
\,\,\,\,\,\,\,\,\, 
[\sqrt{\l}d - (\eta \psi_{+} - \psi_{-})]|_{0,L} = 0
\ee
where $d$ can, in general, be different on  
the two boundaries. 

The action
\be
\label{B.19}
 S = i \int d^2\s\,[\psi_{+}\partial_{-} \psi_{+} + \psi_{-}
 \partial_{+} \psi_{-}] + \frac{i}{2}
 \int d\tau\,[ d\partial_\tau d  + \sqrt{\l}d
 (\eta\psi_{+} + \psi_{-})]|^{L}_{0}
\ee
 reproduces the required equations of motion (\ref{B.16}) and boundary
 conditions (\ref{B.18}). This action is suitable for the $\l = 0$ limit. 
 An equivalent action can be obtained for the $\l = \infty$ limit by 
 absorbing $\sqrt{\l}$ into the definition of $d$. As far as we know, an 
 auxiliary boundary fermion has also been used in the study of the Kondo 
 model with a bulk mass term 
 \cite{Bassi:1999st}.

It is evident from the action (\ref{B.19}) (or, alternatively, from the 
boundary conditions (\ref{B.18})) that in the UV limit ($\l =0$), the
boundary fermion $d$ is needed only for extra zero modes and does not 
play any other role. Such a boundary fermion was also introduced by Witten 
in \cite{Witten:1998cd}, for slightly different reasons . 

As we flow to the IR ($\l \rightarrow \infty$),  $d$ decouples completely. 
We would like to emphasize that this boundary fermion $d$ is just a useful 
tool which facilitates analysis of the theory. All the results we obtain
using $d$ could equally well have been obtained from a direct analysis of
the 
condition (\ref{B.15}).
  
The model given by the action (\ref{B.19}) can be solved explicitly. 
We start by considering the Ramond sector, where the mode
expansion is over integers. For $\l \neq 0$, the properly normalized
solutions
 of (\ref{B.16}) and (\ref{B.18}) are
\be
\label{B.20}
 \psi_{+}(\tau+\s)= \frac{1}{\sqrt{L}}\sum\limits_{r\in Z} 
 \frac{ir+\tilde{\l}}{\sqrt{r^2+\tilde{\l}^2}} \theta_r 
 e^{-ir\frac{\pi}{L}(\tau+\s)},\,\,\,\,\,\,\,\,\,\,\,\,\,\,\,\,
 \psi_{-}(\tau-\s)= \frac{1}{\sqrt{L}}\sum\limits_{r\in Z} 
 \eta \frac{ir-\tilde{\l}}{\sqrt{r^2+\tilde{\l}^2}} \theta_r 
 e^{-ir\frac{\pi}{L}(\tau-\s)} 
\ee
 where the modes obey standard anticommutation relations 
$\{\theta_r, \theta_s\}=\delta_{r+s}$. 

We can now introduce the fermionic currents
\be
\label{B.21}
 j(z)= \sqrt{L}\psi_{+}= \sum\limits_{r\in Z} j_r z^r,\,\,\,\,\,\,\,\,
\,\,\,\,\,
\bar{j}(\bar{z}) =\sqrt{L}\psi_{-} = \sum\limits_{r\in Z} \bar{j}_r
 \bar{z}^r
\ee
 where $z$ and $\bar{z}$ have been defined previously. 
 Our boundary conditions result in the following
 relation between components of currents
\be
\label{B.22}
 j_r - \eta e^{i\varphi(r,\tilde{\l})} \bar{j}_r =
0,\,\,\,\,\,\,\,\,\,\,\,\,\,
  e^{i\varphi(r,\tilde{\l})} = \frac{ir+\tilde{\l}}{ir-\tilde{\l}},
\ee
 where the phase $e^{i\varphi(r,\tilde{\l})}$ has the same limits
 as in (\ref{A.13}). This, together with the
 anticommutation relation $\{j_r,j_s\}=\delta_{r+s}$, leads to the 
following algebra
\be
\label{B.23}
 \{j_r,j_s\}=\delta_{r+s},\,\,\,\,\,\,\,\,\,\,\,\,\,\,\,
  \{\bar{j}_r,\bar{j}_s\}=\delta_{r+s},\,\,\,\,\,\,\,\,\,\,\,\,\,\,\,
 \{j_r,\bar{j}_s\}=\eta e^{i\varphi(s,\tilde{\l})}\delta_{r+s}.
\ee
 Using (\ref{B.23}) one can construct generators which obey the 
 left and right Virasoro algebras. These are given by:
\be
L_k = \frac{1}{2} \sum\limits_{r} ( r + \frac{1}{2} k)
j_{-r} j_{k+r},\,\,\,\,\,\,\,\,\,\,\,\,\,\,
\bar{L}_k = \frac{1}{2} \sum\limits_{r} ( r + \frac{1}{2} k)
\bar{j}_{-r} \bar{j}_{k+r}.
\ee
 Note that $L_k\neq \bar{L}_k$. 
 For $\l\neq 0$ there is just one zero mode $\theta_0$ (hence there 
 is just one ground state)\footnote{This statement applies to Majorana
spinors. For Dirac spinors, one zero mode gives rise to a twice
degenerate vacuum \cite{Ginsparg:1988ui}, because of the algebra 
$\{ \theta_0, \theta_0^{\dagger} \} = 1$,
where $\dagger$ denotes Dirac conjugation.}. 

 At the UV fixed point however, there are 
 two zero modes $\theta_0^+$ and $\theta_0^-$. The canonical commutation 
relations between $\psi_+$ and $\psi_-$ imply that the zero modes obey 
$\{ \theta_0^{\alpha},\theta_0^{\beta}\} = \delta^{\alpha \beta}$. The 
ground state is labelled by a representation of the corresponding 
two-dimensional Clifford algebra. We can define $(-1)^F$ in a natural way, 
 as follows:
\be
\label{B.23A}
 (-1)^F = \theta_0^+ \theta_0^- (-1)^N,\,\,\,\,\,\,\,\,\,\,
 N= \sum\limits_{r} \theta_{-r} \theta_r.
\ee  
The two vacuua then have opposite $(-1)^F$ eigenvalues. World-sheet
parity, 
$\Omega$, plays a remarkably similar role. On the modes, it acts 
 as follows:
 $$ \Omega \theta_0^+ \Omega^{-1} = \theta_0^-$$
 $$ \Omega \theta_0^- \Omega^{-1} = \theta_0^+$$
\be
\Omega \theta_r \Omega^{-1} = e^{i \pi r} \theta_r
\ee
where we have used 
$\Omega \psi_+ \Omega^{-1} = \psi_-$ and $\Omega \psi_- \Omega^{-1} =
\psi_+$. 

Since $\Omega$
interchanges $\theta_0^+$ and $\theta_0^-$, the two vacuua have opposite 
eigenvalues under parity. There is ofcourse an overall ambiguity as to 
which of the two vacuums we choose to be odd and which to be even under 
parity. We can fix this freedom be requiring that each of the two vacua 
should have identical eigenvalues under both $\Omega$ and $(-1)^F$. 
By interchanging the two zero modes, parity also changes the
value of $(-1)^F$. 

It is rather interesting to note that since $\Omega$ and $(-1)^F$ perform
identical functions, we can also obtain the standard Ramond sector by 
moding out the Ramond sector in our theory, by worldsheet parity $\Omega$. 
Thus, at $\l = 0$, making a GSO projection is 'equivalent' to
orientifolding the model by $\Omega$.

>From (\ref{B.15}), it is clear that for $\l \neq 0$, a worldsheet parity 
transformation is not a symmetry and is infact  
equivalent to a change in the sign of $\l$. At the level of the action 
one can see this by rescaling $d$ by $\sqrt{\l}$ 
\be
\label{B.24}
 S = i \int d^2\s\,[\psi_{+}\partial_{-} \psi_{+} + \psi_{-}
 \partial_{+} \psi_{-}] + \frac{i}{2}
 \int d\tau\,[ \frac{1}{\l}d\partial_\tau d  + d
 (\eta\psi_{+} + \psi_{-})]|^{L}_{0}.
\ee 
At the UV fixed point, $\Omega: \sigma \rightarrow L-\sigma$ takes 
$\l$ to $-\l$ and $d$ to  $-d$. 
Due to the boundary fermions (or two fold degeneracy of the ground state), 
the standard Ramond sector (with
only one vacuum) can be thought of as the GSO projected Ramond 
sector, as obtained at the UV fixed point, of our model. This picture 
is similar to what Witten has proposed in \cite{Witten:1998cd}. When $\l$ 
becomes non zero the model picks up one of the two ground states and flows
to 
the IR. In this sense one can say that the RG flow is 'GSO projected' with 
respect to the UV fixed  point. Which particular ground state is projected
out
depends on the sign of $\l$.

The NS sector can also be treated in a similar fashion. 
For  $\l\neq 0$, the equations (\ref{B.20})-(\ref{B.23}) carry through to
the NS sector, with the modification that $r$ now takes half integer
values. 
However, things get a little more complicated at the UV point fixed point. 
Once again (as in the R sector), the vacuum has a two fold degeneracy.
This
degeneracy can be attributed to the two extra boundary fermions that exist
at 
the two boundaries. Since these fermions do not enter into the
Hamiltonian, 
they do not affect the energy of the system; all they do is provide a 
mechanism for making the vacuum degenerate.  
In this sector, we define the action of parity\footnote{
At the level of the action, we have the freedom of defining worldsheet
partity 
$\Omega$ such that $\Omega^2 = 1$ or $-1$. We decide which alternative is
more
suitable by appealing to the boundary conditions. In the R sector, we 
find that boundary conditions are preserved if we choose $\Omega^2 = 1$
while
in the NS sector, boundary conditions are parity invariant when parity is 
defined such that $\Omega^2 = -1$ \cite{GP:1996od}.} as follows:
$\Omega \psi_+ \Omega^{-1} = - \psi_-$ and 
$\Omega \psi_- \Omega^{-1} = \psi_+$. This implies that 
$$ \Omega \theta_0^+ \Omega^{-1} = - \theta_0^-$$
$$ \Omega \theta_0^- \Omega^{-1} = \theta_0^+$$
\be
\Omega \theta_r \Omega^{-1} = e^{i \pi r} \theta_r
\ee
With $\Omega$ defined thus, and $(-1)^F$ defined as in (\ref{B.23A}), we
can 
now proceed just as we did in the Ramond sector.

 So far our arguments were based on the canonical quantization approach.
 We were trying to quantize the theory using the bulk equations of motion
 (\ref{B.16}) and the boundary conditions (\ref{B.15}), so the action
 (\ref{B.19}) was useful but was not necessary. However the action becomes
 a crucial tool when one makes a Wick rotation and goes to the Euclidean 
 version; for (\ref{B.19}), this has the following form
\be
\label{B.26}
 S= i \int\limits_{\Sigma} d^2z[ \psi \partial_{\bar{z}}\psi +
 \bar{\psi} \partial_{z} \bar{\psi}] - 
 \frac{i}{2} \int\limits_{\partial \Sigma} [d\partial_t d +
 \sqrt{\l} d(\psi + \bar{\psi})]
\ee  
 where $\Sigma$ is some domain in $R^2$ and $t$ parameterizes the boundary 
 of $\Sigma$. 
 Starting from (\ref{B.19}) we made 
 the Wick rotation ($\tau =it$) and redefined the spinors
 $\psi= z^{1/2}\psi_{+}$ and $\bar{\psi}=\bar{z}^{1/2}\psi_-$ 
 (from now on $\psi_+$ and $\psi_-$
  should be thought of as Weyl spinors which are 
 complex conjugates of each other). $d$ is a real boundary
 fermion (equivalently, it can be made purely imaginary). 
 Fermions are periodic or antiperiodic depending on whether they
 belong to the R or NS sector. This model can now be treated using
standard 
 path integral techniques, where integration over $d$ is assumed. 
 
 \section{Supersymmetric theory}

 Based on results from the two previous sections we can now construct 
 the supersymmetric version of our theory. The supersymmetric action is
 simply the sum of (\ref{A.1}) and (\ref{B.19}) 
$$ S= \frac{1}{2}\int d^2\s [\partial_{-}\phi \partial_{+} \phi
 + i \psi_{+} \partial_{-} \psi_{+} + i \psi_{-} \partial_{+} \psi_{-}]
  - \frac{1}{2} \int d\tau \l \phi^2\,|^L_0 $$
\be
\label{C.1} 
+ \frac{i}{4} \int d\tau [d\partial_\tau d + \sqrt{\l}d(\eta \psi_{+}
 + \psi_{-})]|^L_0.
\ee
 This action gives rise to the boundary conditions (\ref{A.6})
 and (\ref{B.18}). The presence of a boundary introduces subtleties
with supersymmetry, hence, together with bulk supersymmetry
transformations,
we need to use the above boundary conditions as well, when checking for
supersymmetry \cite{Haggi-Mani:2000uc}. Using the bulk transformations 
\be
\label{C.2}
 \delta \phi = - \epsilon^{+} \psi_{+} - \epsilon^{-} \psi_{-}, 
\,\,\,\,\,\,\,\,\,\,\,\,\,\,\,
\delta \psi_{+} = -i\epsilon^+ \partial_{+}\phi,
\,\,\,\,\,\,\,\,\,\,\,\,\,\,\,
\delta \psi_{-} = - i\epsilon^{-} \partial_{-} \phi,
\ee
 together with the transformation of the boundary fermion
\be
\label{C.3}
\delta d = 2i\epsilon \sqrt{\l} \phi,
\ee  
 one can show that the supersymetric variation of the action (\ref{C.1})
 is zero, modulo boundary conditions (\ref{A.6}) and (\ref{B.18}). 
 This action is similar to the component form action proposed in 
 \cite{Harvey:2000na} for the case of a quadratic tachyonic potential. 
 However it seems that in general, the proposed action does not have the 
 right properties.  
 The superfield formalism does not guarantee supersymmetry in the presence 
 of a boundary, unless boundary conditions are suitably taken into
account, 
 in accordance with procedures outlines in \cite{Haggi-Mani:2000uc}.

 So far we have been unable to construct the supersymmetric action for a
 general potential $V(\phi)$. However, the general boundary conditions 
 (\ref{A.2}) and (\ref{B.17}) are compatible with supersymmetry and 
 ensure closure of the supersymetric algebra classically. 

 The supersymmetric model with a quadratic potential can be worked out 
 explicitly. Starting from the 
Ramond sector, using the solutions (\ref{A.7}) and (\ref{B.20}) we can
define
the components of the currents $J_n$ and $j_r$  as in (\ref{A.8}) and 
 in (\ref{B.21}) correspondingly. We can also construct generators
\be
\label{C.4}
 L_k = \frac{1}{2}\sum\limits_{n}J_n J_{k-n} + \frac{1}{2} \sum\limits_{r}
 (r+\frac{1}{2}k)j_{-r} j_{k+r},
\,\,\,\,\,\,\,\,\,\,\,\,\,\,\,\,\,\,\,\,\,\,\,
G_k= \sum\limits_{n} J_{-n} j_{k+n}
\ee   
 which obey the super Virasoro algebra. In the same way, generators 
 $\bar{L}_k$ and $\bar{G}_k$ (which obey the same algebra) can be
constructed 
 using the antiholomorphic currents $\bar{J}_n$ and $\bar{j}_n$. 
 For $\tilde{\l}\neq 0$ one can see that 
 $L_k \neq \bar{L}_k$ and $G_k\neq \bar{G}_k$. This is at it should be, as 
 super conformal invarince is broken by boundary interations. 
 At the IR fixed point ($\tilde{\l}= \infty$), the left and right super 
 Virasoro algebras coincide. At the UV fixed point we have the 
 extended super Virasoro algebra \cite{Cohn:1988hw} (i.e., super Virasoro 
 algebra together with $(-1)^F$). 

 An identical analysis can be performed for the NS sector with the only 
 difference that now we have half integer moding for fermionic currents 
 $j_r$, $\bar{j}_r$ and generators $G_k$, $\bar{G}_k$. At the UV fixed
 point, due to the doubly degenerate vacuum, we would have two copies of
the representation of NS super Virasoro algebra.   
 
One may wonder about the need of introducing a degeneracy into the vacuum
at
the UV fixed point. There is a simple explanation. Suppose that in the
action 
(\ref{C.1}) we replace the bosonic boundary potential by the following 
expression 
\be
- \frac{1}{2} \lambda \int\ d\tau (\phi - \phi_1)^2 |_L + 
 \frac{1}{2} \lambda \int\ d\tau (\phi - \phi_0)^2 |_0. 
\ee
The action thus modified will still be supersymmetric. Depending on the 
sign of $\lambda$, we will flow, in the IR to different theories. One of
these 
will have $\phi|_0 = \phi_0, \;\;\;\; \phi|_L = \phi_1$ and the other, 
$\phi|_0 = \phi_1, \;\;\;\; \phi|_L = \phi_0$. These theories are
related by a parity transformation $\Omega$ and they each have their own 
vacuum. Therefore, inorder to be able to get either of these theories in
the 
IR, we need to have both the corresponding vacuua present in the theory at 
the UV fixed point.

\section{Non-trivial realization of bulk symmetries}

 In this section we would like to investigate the possibility of having
 different perturbations on the two boundaries. We start by 
 considering the bosonic model given by the following action
\be
\label{D.1}
L= \frac{1}{2} \int\limits_{0}^{L} d\s\, \d_\alpha \phi \d^\alpha \phi + 
 \lambda V(\phi(0)) - \gamma V(\phi(L)) .
\ee 
 This model has a rich phase structure with four fixed points NN, ND, DN
and
 DD, where N and D stand for Neumann and Dirichlet boundary conditions 
 respectively. Obviously, our previous bosonic model can be easily
embedded 
 into this new one. Also, when $\gamma =-\l$, the model becomes parity 
 invariant, unlike previous examples. 

 As before, we can work out the quadratic potential 
 $V(\phi)= \frac{1}{2} \phi^2$ in detail. 
 In this case we get the following boundary conditions
\be
\label{D.2}
 (\phi' + \l \phi)|_{0}=0,\,\,\,\,\,\,\,\,\,\,\,\,\,
 (\phi'+\gamma \phi) |_{L} =0.
\ee
 The solution of the  massless Klein-Gordon equation with boundary 
 conditions (\ref{D.2}) can also be reduced to a Sturm-Liouville
 problem, the normalized solution of which is given by the following 
expresion
\be
\label{D.3}
\phi(\tau,\sigma) = \frac{\sqrt{2L}}{\pi} \sum\limits_{\tilde{n}\neq 0}
 \frac{e^{-i{\tilde{n}}\frac{\pi}{L} \tau}}{\sqrt{\tilde{n}^2 
 + \tilde{\lambda}^2}}
 a_{\tilde{n}} [\cos(\tilde{n}\frac{\pi}{L}\sigma) - 
 \frac{\tilde{\lambda}}{\tilde{n}}
 \sin(\tilde{n}\frac{\pi}{L}\sigma)], 
\ee
 where $\tilde{n}$ is subject to the transcendental equation
\be
\label{D.4}
 \tan (\tilde{n}\pi) = \frac{(\tilde{\gamma}-\tilde{\lambda})\tilde{n}}
 {\tilde{n}^2+\tilde{\gamma}\tilde{\lambda}}
\ee
 with $\tilde{\lambda}= \frac{\lambda L}{\pi}$ and
 $\tilde{\gamma} = \frac{\gamma L}{\pi}$. Equation (\ref{D.4}) cannot be 
 solved explicitly, but positive solutions $\tilde{n}(\tilde{\l},
 \tilde{\gamma})$ can be ordered and presented in the following form
\be
\label{D.5}
 \tilde{n}(\tilde{\l},\tilde{\gamma})= n + g(\tilde{\l},
\tilde{\gamma},n),
\,\,\,\,\,\,\,\,\,\,\,\,n\in Z_{+},\,\,\,\,0\leq 
g(\tilde{\l}, \tilde{\gamma},n) < 1 
\ee
 where $g$ is a function of $\tilde{\l}$, 
 $\tilde{\gamma}$ and $n$ which can be computed numerically. Negative
 solutions are obtained trivially by taking 
 $\tilde{n} \rightarrow - \tilde{n}$. 
 The asymptotic
 behaviour of $g$ is clear; it must be zero at the fixed points 
 ($\tilde{\l}=0,\tilde{\gamma}=0$), 
 ($\tilde{\l}=\infty, \tilde{\gamma}=\infty$) 
 and $1/2$ at the fixed points 
 ($\tilde{\l}=\infty,\tilde{\gamma}=0$), 
 ($\tilde{\l}=0,\tilde{\gamma}=\infty$).
 
 We can now relabel $a_{\tilde{n}}$ as $a_n$ and define 
 chiral and antichiral currents which have the following
 components
\be
\label{D.6}
 J_n = -\frac{\tilde{\l}+i\tilde{n}}{\sqrt{\tilde{n}^2+\tilde{\l}^2}}
 e^{-i\frac{\pi}{L}(\tau +\s)g(\tilde{\l},\tilde{\gamma},n)} a_n,
\,\,\,\,\,\,\,\,\,\,\,\,\,\,\,
 \bar{J}_n = \frac{\tilde{\l}-i\tilde{n}}{\sqrt{\tilde{n}^2+\tilde{\l}^2}}
 e^{-i\frac{\pi}{L}(\tau -\s)g(\tilde{\l},\tilde{\gamma},n)} a_n.
\ee
 Thus the components of chiral and antichiral currents are related as
follows:
\be
\label{D.7}
 J_n - e^{i\varphi(n,\tilde{\l},\tilde{\gamma})} \bar{J}_n = 0,
\,\,\,\,\,\,\,\,\,\,\,\,\,\, e^{i\varphi(n,\tilde{\l},\tilde{\gamma})}=
 \frac{i\tilde{n}+\tilde{\l}}{i\tilde{n}-\tilde{\l}} e^{-2i\frac{\pi}{L}\s 
 g(\tilde{\l}, \tilde{\gamma},n)}.
\ee
 Using canonical commutation relations, together with the fact that 
 eigenfunctions of a Sturm-Liouville problem form a complete set,
 one can show that the currents (\ref{D.6}) obey the same algebra as 
 (\ref{A.10}) but with a new
 spatial dependent phase  $e^{i\varphi(n,\tilde{\l},\tilde{\gamma})}$.

 As before, using the Sugawara construction (\ref{A.7}) we construct
 left and right Virasoro generators $L_n$ and $\bar{L}_n$ such that
 $L_n\neq \bar{L}_n$ at a generic point in the 
 $(\tilde{\l},\tilde{\gamma})$-plane. At the fixed points 
 ($\tilde{\l}=0,\infty;\tilde{\gamma}=0,\infty$) conformal
 symmetry is restored and we have $L_n= \bar{L}_n$. 
 In the present model we see that bulk conformal symmetry is realised
rather
 non-trivially. The representation of the left Virasoro algebra has a 
 natural realization on the Fock space built up by $J_n$ whereas the 
 representation of the right Virasoro algebra would be realized naturally 
 on the Fock space built up by $\bar{J}_n$. These two Fock spaces are 
 related through a spatially dependent unitary transformation (\ref{D.7}).  
 Hence, $\sigma$-independence for $L_n$ in a particular basis would
 necessarily imply $\s$ dependence for $\bar{L}_n$. 

 In general, the model (\ref{D.1}) is not parity invariant. A parity 
 transformation $\Omega$ amounts to changing the signs of the coupling 
 constants: $(\l,\gamma) \rightarrow (-\l,-\gamma)$. However, there is a 
 fixed line $\gamma= -\l$ under this transformation where the model
becomes 
 parity invariant. For a model defined on this line, parity symmetry would
be
 realized in a highly nontrivial fashion since its action on the Fock
space would be $\sigma$-dependent. At the self-dual point
$\sigma=L/2$,
 the left and right Virasoro generators coincide $L_n=\bar{L}_n$. It is
not
 clear to us how to interpret this result. 
  
 The fermionic model can also be easily generalized in the same fashion. 
 The most general scenario for fermions is given by the following
 boundary conditions 
$$[\d_\tau(\eta_0 \psi_+ - \psi_-) + \lambda (\eta_0 \psi_+ +
\psi_-)]|_{0}=0$$
\be
\label{D.8}
[\d_\tau(\eta_L \psi_+ - \psi_-) + \gamma (\eta_L \psi_+ + \psi_-)]|_{L}=0
\ee
 where $(\eta_0,\eta_L)$ corresponds to the choice of spin structure at
 the two boundaries. 
 The general solutions to this problem are given below
\be
\label{D.9}
 \psi_{+}(\tau+\s)= \frac{1}{\sqrt{L}}\sum\limits_{\tilde{r}} 
 \frac{i\tilde{r}+\tilde{\l}}{\sqrt{\tilde{r}^2+\tilde{\l}^2}} 
 \theta_{\tilde{r}} 
 e^{-i\tilde{r}\frac{\pi}{L}(\tau+\s)},\,\,\,\,\,\,\,\,\,\,\,\,\,\,\,\,
 \psi_{-}(\tau-\s)= \frac{1}{\sqrt{L}}\sum\limits_{\tilde{r}} 
 \eta_0 \frac{i\tilde{r}-\tilde{\l}}{\sqrt{\tilde{r}^2+\tilde{\l}^2}} 
 \theta_{\tilde{r}} 
 e^{-i\tilde{r}\frac{\pi}{L}(\tau-\s)} 
\ee
 where $\tilde{r}$ is subject to the constraints
\be
\label{D.10}
 \tan(\tilde{r}\pi) =  \frac{(\tilde{\gamma}-\tilde{\lambda})\tilde{r}}
 {\tilde{r}^2+\tilde{\gamma}\tilde{\lambda}},\,\,\,\,\,\,\,\,\,\,\,\,\,\,
\,\,\,\,\,\,\,\,for\,\,\,\,\,\eta_0\eta_L=1,
\ee
\be
\label{D.11}
 \tan(\tilde{r}\pi) =  \frac{\tilde{r}^2+\tilde{\gamma}\tilde{\lambda}}
{(\tilde{\gamma}-\tilde{\lambda})\tilde{r}},\,\,\,\,\,\,\,\,\,\,\,\,\,\,
\,\,\,\,\,\,\,\,for\,\,\,\,\,\eta_0\eta_L = -1.
\ee
 Thus, depending on the sector, $\tilde{r}$ obeys different transcendental 
 equations. In the sector $\eta_0\eta_L=1$ the condition (\ref{D.10}) for 
 fermions coincides with the bosonic condition (\ref{D.4}), so fermionic
 currents $j_r$ and $\bar{j}_r$ can be constructed just as in the bosonic
case.
 These currents are related as follows:
\be
\label{D.12}
 j_r - e^{i\varphi(n,\tilde{\l},\tilde{\gamma})} \bar{j}_r =0,
\ee 
 where the spatial dependent phase is given by second equation in
(\ref{D.7}).
 It is fairly straightforward to construct the left and right super
Virasoro 
 algebras. Their realizations on the Fock space are $\sigma$-dependent, 
 exactly as for the bosonic model. 

 For the sector $\eta_0\eta_L=-1$, fermionic and bosonic modings are
 different. This is not surprising since, even at the fixed points, 
 the moding is different for bosons and NS fermions. We work around this 
 problem by using the same tricks as in the bosonic case.
 It is natural now to write $\tilde{r}$ as the sum of a function and a
half 
 integer; this makes the $\s$-dependent phase between $j_r$ and
$\bar{j}_r$
 different to what it was before (\ref{D.7}). The left and right super 
 Virasoro algebras can be constructed in the usual way.   

 So far we have considered the theory away from critical points. As
before, 
 the UV fixed point needs special consideration due to the extra zero 
 modes which arise there. Once again, there is a natural $(-1)^F$-operator 
 which defines the GSO projection and the RG flow can be thought of as 
 'GSO projected' from the UV point of view, since the degeneracy of the
ground 
 state is lifted. In the present model, parity does not 
 in general play the role it played in section 3. Away from the line 
 $\l= - \gamma$ we can use parity the way we used it earlier 
 and argue that the change $(\l,\gamma) \rightarrow (-\l,-\gamma)$ leads 
 to a change in the value of $(-1)^F$; at the line $\l=-\gamma$
 where parity is realized non trivially, we cannot use this 
 argument. 

\section{String theory applications}

 The main string theory applications of our results come from the
 boundary renormalization group interpretation of open string 
 tachyon condensation \cite{Harvey:2000na}. Formally one can rewrite 
 the action (\ref{C.1}) for a number of bosons and fermions,
 labelled by an external space time index: $X^\mu = \sqrt{2\pi\alpha'}
 \phi^\mu$ and $\Psi_{\pm}^\mu = \sqrt{2\pi\alpha'}\psi_{\pm}^\mu$, 
 where $X$ and $\Psi$ are canonically normalized. 
 The boundary potential, called the tachyon
 profile henceforth, can now depend on more than one direction.
 Various string models arise, resulting from the freedom of 
 assigning different boundary conditions. The analysis of such models
should
 be similar to what we have outlined in this paper for a more simple case.  
 Even without studying these models in detail however, we can find certain
 restrictions they need to obey. 

 Depending on the details of the string configuration we choose and 
 the number of directions on which the tachyonic profile depends, 
 we get an even or odd dimensional Clifford algebra for zero modes at 
 the UV fixed point. It is obvious that the operator $(-1)^F$ 
 can be defined meaningfully only for an even dimensional Clifford 
 algebra \cite{Witten:1998cd}. Thus, it is only these cases to which we
can 
 hope our general arguments will apply. This may be interpreted 
 as a restriction on the D-brane systems which can consistently couple 
 to a particular tachyonic profile. 

 Another problem is how to properly interpret the UV fixed point of this
 model within the framework of string theory i.e, finding a  
 string theory interpretation of the auxilary fermions living on the
boundary.
 These boundary fermions increase the degeneracy of the ground state,
having
 much the same effect as Chan-Paton factors. Keeping this analogy in mind,
we 
 suppose that the UV fixed point should correspond not to a single
standard 
 open string sector but instead to a few sectors, the particulars of which 
 are determined by the algebra of the boundary fermions. 

 We would also like to point out the following fact: Inorder
 to couple an open string to a tachyonic background, we need to introduce 
 boundary fermions and through them, a number of degenerate open string 
 vacuua. Building on the analogy with Chan Paton factors, it would seem
that 
 these open string vaccua interpolate between different D-brane
configurations; i.e, that it is not consistent to couple a background
tachyon to an open 
 string living on a single D-brane.

There are several obvious directions for further work. To begin with, it 
would be worthwhile to write down the supersymmetric action for a general 
tachyonic background; we have been able to do so only in the case when the 
tachyonic profile is quadractic. Also, while it is clear that our
arguments 
can only be expected to work for even dimensional Clifford algebras,
explicit 
examples of such cases have not yet been looked at. Concrete
configurations 
of D-branes can be studied to work out the nature of the restrictions on
the 
tachyonic profiles which may consistently couple to it, and indeed to
verify 
that such restrictions even exist. Similarly, the correspondence between
the
degeneracy introduced by Chan-Paton factors and that introduced by
boundary
fermions should be explored in more detail and the relation made precise. 
This, again, should tell us exactly which configurations of D-branes need
to
be present inorder for us to couple the open strings to a tachyonic
profile. 
Ofcourse we would expect the allowed D-brane configurations resulting from 
both the above analyses to be identical. 
 
 \begin{flushleft} {\Large\bf Acknowledgments} \end{flushleft}

 We are grateful to Hans Hansson for clarifying many points and to 
 Ansar Fayyazuddin, Daniel Lillieh\"o\"ok,
 Ulf Lindstr\"om and Subir Mukhopadhyay for useful discussions. 

%\newpage
\noindent

%**********************************

\end{document}